\documentclass[twocolumn]{aastex631}

\newcommand{\msun}{\;\mathrm{M_\odot}}

\usepackage{amsmath}
\usepackage{rotating}
\usepackage{makecell}
\usepackage{booktabs}

\begin{document}

\title{Massive interacting binaries as an enrichment source for multiple populations in star clusters}

\correspondingauthor{Alison Sills}
\email{asills@mcmaster.ca}

\author{Michelle Nguyen}
\affiliation{Department of Physics \& Astronomy, McMaster University \\
1280 Main Street West \\
Hamilton, L8S 4M1, Canada}

\author[0000-0003-3551-5090]{Alison Sills}
\affiliation{Department of Physics \& Astronomy, McMaster University \\
1280 Main Street West \\
Hamilton, L8S 4M1, Canada}

\begin{abstract}

We present a suite of binary evolution models with massive primaries ($10\leq M_1 \leq 40 \msun$) and periods and mass ratios chosen such that the systems undergo non-conservative mass transfer while the primaries have helium cores. We track the total mass and chemical composition of the ejecta from these systems. This material shows the abundance signatures of hot hydrogen burning which are needed to explain the abundance patterns seen in multiple populations in massive star clusters.  We then calculate the total yield of a population of binary stars with masses, mass ratios, and periods consistent with their distribution in a field population. We show that the overall abundance of this material is enriched in helium, nitrogen, sodium, and aluminum, and depleted in carbon, oxygen, and magnesium, by amounts that are consistent with observations. We also show that such a population of binaries will return approximately 25\% of its mass in this ejecta (compared to 4\% if all the stars were single), over a characteristic timescale of about 12 Myr. We argue that massive binaries must be seriously considered as a contributor to the source of enriched material needed to explain the multiple populations in massive clusters, since essentially all massive stars are formed in binaries or higher order multiples, massive binaries are primarily formed in clusters, and massive binaries naturally produce material of the right composition. 

\end{abstract}


\section{Introduction}

Massive star clusters were for many years thought to be simple stellar populations, consisting of stars formed at the same time, and out of material with a uniform composition. This assumption is still reasonable for the purposes of determining their ages or metallicities \citep[e.g.][]{2023MNRAS.522.5320D,2023MNRAS.525L...6A,2023AJ....166...18Y}. As simple populations, clusters can encode the galactic conditions at the time and place when the cluster formed, and therefore can trace galaxy formation and assembly \citep{2020SSRv..216...64K}. Clusters have also been invaluable for our understanding of stellar evolution and stellar dynamics, since they were thought to provide constraints such as a common age, composition, and distance which are usually unavailable for individual stars.

However, the advent of exquisite Hubble Space Telescope photometry of large number of globular clusters in the Milky Way, and accompanying high resolution spectroscopy, confirmed the hints from many decades earlier that in fact cluster stars do not have uniform surface abundances. While many stars show the same chemical patterns as field stars of the same metallicity, approximately half of the stars show variations. In particular, we see correlations or anti-correlations in light elements but (generally) no variation in iron-peak or s-process elements. Specifically, the enriched population shows enhancement in helium, nitrogen, sodium; depletion in carbon and oxygen; and in a few clusters there is an increase in aluminum for either constant or decreased magnesium. The variations are consistent with material that has been processed by the nucleosynthetic cycles associated with hot hydrogen burning beyond the CNO cycle (at about 70 MK) but not subsequent burning. Essentially all massive clusters show evidence of these so-called ``multiple populations" when studied carefully, regardless of cluster age or galactic environment. For recent reviews of the state of the observational evidence for multiple populations in globular clusters, see \citet{2019A&ARv..27....8G,mult,2022Univ....8..359M}. 

The origin of these multiple populations is still not clear. Any complete model must identify a nucleosynthetic site capable of processing hydrogen-rich material at ~70 MK. That material must then be made available to either form new stars or be added to the surfaces of pre-existing stars. And finally there must be enough material available so that half or more of the present-day stars in a cluster have the modified abundances. The nucleosynthetic source must also primarily act in clusters, as only a small percentage of field stars show these same chemical patterns \citep{2011A&A...534A.136M}. It must act on a timescale that is quite short, as there is no demonstrable age difference between the various populations. In addition, the abundance anomalies are found in stars throughout the HR diagram, suggesting that the stars were formed with this composition rather than being polluted later in their lives.

There are handful of viable sources for the material with the appropriate composition (usually called `enriched' material even though some elements are depleted). These include asymptotic giant branch stars \citep{2009A&A...499..835V}, rapidly-rotating massive stars \citep{frms}, very massive ($\sim300 \msun$) or supermassive \citep[$\gtrsim 5000 \msun$][]{2018MNRAS.478.2461G} stars, stellar collisions \citep{2010MNRAS.407..277S}, and interacting massive binary stars \citep{deMink2009}. While many of these sources have been explored in detail (particularly asymptotic giant branch stars and rotating massive stars), the massive binary stars have been given less attention. The original paper that proposed these objects as a source of material \citet{deMink2009} modelled only one binary system, essentially as a proof of concept. 

However, massive binaries are an attractive solution to the multiple population problem. Multiple populations are found in essentially every massive cluster studied, and so the mechanism that creates them must be a normal outcome of star formation. Massive stars are preferentially found in binary or higher order systems, with a multiplicity fraction of 100\% for stars more massive than $\sim10 \msun$  \citep{2023ASPC..534..275O}. Binary interactions (e.g. mass transfer, mergers, and common envelope evolution) dominate the evolution of massive stars \citep{2012Sci...337..444S}. In dense regions, dynamical evolution of binaries will primarily drive them to shorter orbital periods, potentially increasing the number of interacting massive binaries in clusters. In a previous study, we have demonstrated the viability of a pollution source which is proportional to the number of massive stars and which acts during the cluster assembly process inside the cluster's natal giant molecular cloud \citep{2019MNRAS.486.1146H}. 

 The binary system investigated by \citet{deMink2009} consisted of two massive stars with an orbital period such that mass transfer would occur after core hydrogen exhaustion but before core helium burning had begun. They demonstrated that, with an appropriate treatment of angular momentum which spins up the secondary and drives non-conservative mass transfer, significant amounts of processed material could be ejected from the system. In this paper,  we revisit massive interacting binary stars as a pollution source for multiple populations in massive clusters. We present a suite of binary evolution models covering a larger range of masses and orbit properties, and quantify the yields from these systems. We then look at the ensemble yield from a population of such binaries. Finally, we comment on the viability of massive interacting binaries as a source of enriched material which should be considered in a solution to the multiple population problem.  

\section{Methods}\label{sec:methods}

\subsection{Binary evolution calculations}
\label{sec:physics} 

We simultaneously evolve both components of our binary systems using Modules for Experiments in Stellar Astrophysics (MESA) version r15140 \citep{Paxton2011,Paxton2013,Paxton2015,Paxton2018,Paxton2019,jermyn2023}. We generally used the suggested parameters from the MESA Isochrones and Stellar Tracks (MIST) suite for single star models \citep{Choi2016} and followed the example of \citet{Paxton2015} for binary evolution. A full description of the parameters used can be found in \citet{NguyenThesis}, and sample 
\texttt{inlists}, the custom reaction rate network file, and our \texttt{run\_stars\_extras.f90} \&\texttt{run\_binary\_extras.f90} files are available at \dataset[10.5281/zenodo.10998057]{\doi{10.5281/zenodo.10998057}}. 

We constructed zero-age main sequence (ZAMS) models from the pre-main sequence with an initial alpha-enhanced metallicity of [Fe/H] $=-1.44$, similar to that used by \citet{deMink2009} and in the middle of the range of metallicities of Milky Way globular clusters. We follow the prescription of \citet{reddy2006} for the alpha-enhancement, which results in [O/Fe]=0.44, Z=$1.2\times10^{-3}$ and Y=0.25. These ZAMS models were evolved as single stars to allow us to determine appropriate periods for our binary systems, and as a comparison population for the nucleosynthetic yields. The ZAMS models also serve as the starting point for each of our binary evolution calculations.  Mixing mechanisms appropriate for massive stars are included (convection, semi-convection, overshoot, thermohaline mixing, rotational mixing). Wind mass loss is treated using the `Dutch' scheme, as outlined in \citet{dutch1}. Because we wish to evaluate the nucleosynthetic yields relevant for the multiple population problem, we use a custom reaction network which is an expansion of MESA's `h\_burn.net' and includes a more complete coverage of H, He, and C burning processes including the high-temperature hydrogen burning cycles. At each time step we compute and output the ejection of 50 isotopes of elements up to the iron peak. 

Our choices for binary evolution generally follow those given in \citet{Paxton2015}. We use the default \citet{Ritter1988} mass transfer scheme, which was also used in \citet{deMink2009}. Our binaries are initialized on circular orbits and given an initial rotation rate as set by tidal synchronization. As the binary evolves, angular momentum loss from the orbit is computed based on mass loss unaccounted for by the stellar angular momentum loss. We also allow angular momentum transfer through accretion, both of which are necessary for following the non-conservative mass transfer which takes place in these systems. The maximum rotation rate that a star can have is set to 90\% of its critical rotation rate, as MESA is not numerically stable beyond that rate \citep{Paxton2019}.   

\subsection{Initial Conditions}

For this study, we aim to cover the full parameter space which may produce ejecta with appropriate abundances for the multiple populations problem, while reducing the time spent on simulations which do not add much material to the population. We chose our primary masses $M_1$ to be in the range of $10-40\msun$ since these stars are massive enough to reach high central temperatures. Higher mass stars also reach appropriate temperatures, but as will be described in more detail below, binaries with more massive primaries are unlikely to contribute significant amounts of material of the appropriate abundances for the multiple populations problem.  Mass ratios $q=M_2/M_1$ were evenly sampled over the range of 0.15 - 0.9. This excludes twins, $q\gtrsim0.95$, which if close enough to initiate mass transfer in our simulations, would become contact binaries. It also excludes very unequal mass systems, which behave more and more like single stars or conservatively-transferring systems. 

\begin{figure}
    \centering
    \includegraphics[width=\columnwidth]{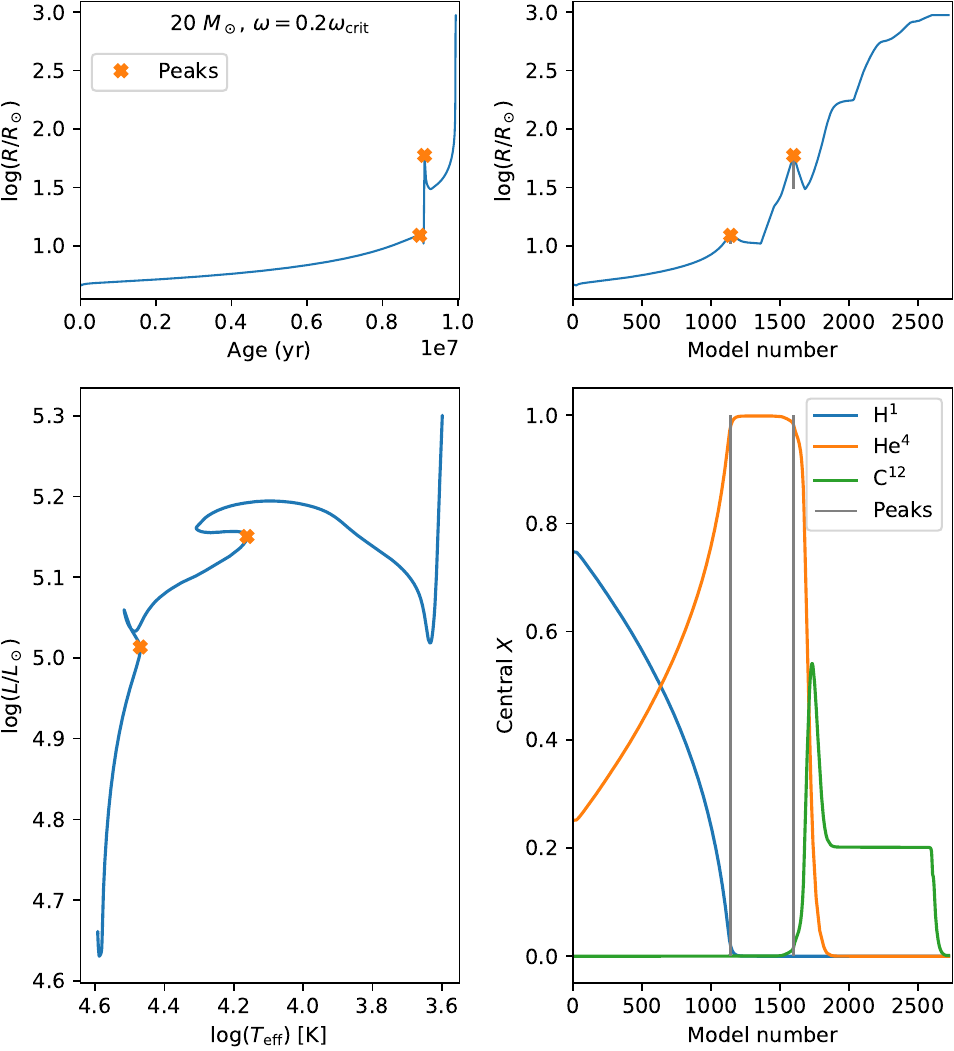}
    \caption{Evolution of an example single model ($20\msun)$, showing the radius evolution in the upper panels, the stellar track in the lower left, and the central abundances in the lower right. The x axis on the right-hand side shows the model number, which increases with time but shows the interesting phases of evolution more clearly. Orange crosses and grey vertical lines correspond to the identified peaks in radius which determine the minimum and maximum periods we allow. }
    \label{fig:single_ex}
\end{figure}

\begin{figure}
	\includegraphics[width=\columnwidth]{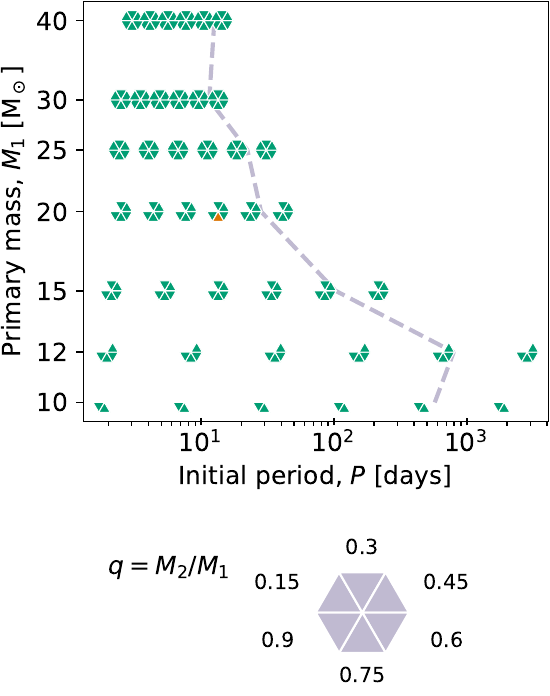}
    \caption{Periods, primary masses, and mass ratios of the binaries simulated in this work.
    Binaries of the same initial primary mass and initial orbital period are triangular segments of a hexagon centred on the primary mass and period.
    The relative orientation of the segment represents the mass ratio, $q$, shown in the legend.
    The grey dashed line represents the expected upper limit on the period for binaries of interest based on single-star models at the sampled primary masses. The orange triangle indicates the system modelled by \citet{deMink2009}.
    }
    \label{fig:param_space}
\end{figure}

Our binary periods were chosen such that the systems would undergo mass transfer between core hydrogen exhaustion and core helium ignition (i.e. case B mass transfer). Using our single-star models, we determined the radii of each of our stars at these two evolutionary stages, which turns to correspond to two peaks in the radius evolution, as shown in Figure \ref{fig:single_ex}. We then set these radii equal to the Roche lobe radius as given by \citet{Eggleton1983}, assuming a mass ratio of 0.5, and solved for the period given this semi-major axis. We also added one longer period for each primary mass, just to confirm that we had in fact bounded the period appropriately. The evolutionary endpoints correspond to periods ranging from approximately 2-700 days, and the longest period we simulate is $\sim2800$ days. 
The lower limit falls into a narrow range of around 2-3 days for the mass range we cover. The upper period limit varies sharply with mass. With a narrower range of periods for the high-mass primaries expected to act as enrichment sources, and with the relative rarity of high-mass primaries, the high-mass binaries in our sample are not expected to contribute significant enrichment. Figure ~\ref{fig:param_space} shows the parameter space sampled by the binaries, spanning period $P$, primary mass $M_1$, and mass ratio $q$. A total of 204 binaries were simulated. 

\subsection{Stopping Conditions}

\begin{figure}
    \includegraphics[width=\columnwidth]{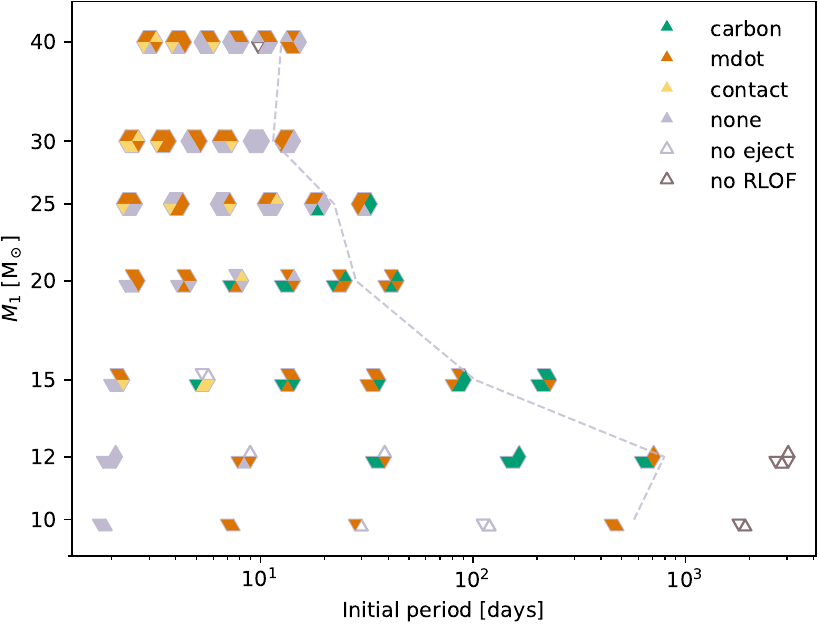}
    \caption{Stopping conditions reached by each binary in the sample of 204 binaries. 
    Filled-in triangles are included in our main sample and empty triangles are omitted. 
    The layout of this figure follows Figure \ref{fig:param_space}.     
    The colour of each point indicates the stopping condition reached by that binary.
    Out of the 190 binaries that we include in our main sample, 29 reach core-C exhaustion, 34 initiate contact, and 67 have sharp changes in their mass loss rates, and 14 out of 204 binaries are excluded for not ejecting at least 0.1~$\msun$.
    }
    \label{fig:stops_alpha}
\end{figure}

Stars within a binary were simultaneously evolved from zero-age main sequence (ZAMS) models until they reached one of our four possible stopping conditions. The endpoint of all 204 of our systems are shown in Figure \ref{fig:stops_alpha}. First, we stopped our simulation when we saw a sudden and very large increase in the system mass loss rate, which signals the onset of dynamically unstable mass transfer and, potentially, the creation of a common envelope system. Second, we ended a run if the stars came into contact, i.e. when both stars are simultaneously overflowing their Roche lobes. These systems usually occur at the shortest periods. MESA has been used to model contact systems for primary  masses below 20 $\msun$ \citep{2024A&A...682A.169H}. In our higher mass models, we encountered both significant numerical instability and uncertainty in the physical validity of different approaches. Therefore, we made the choice to stop our simulations, knowing that we may be under-estimating the amount of material that could be ejected from these systems.

Third, some systems evolved to core carbon depletion, which we defined as the time when the central $^{12}$C mass fraction drops to $5\times10^{-3}$ along with a low central He mass fraction ($<1\times10^{-4}$). These are typically the intermediate mass primaries ($12\lesssim M_1\lesssim25\;\mathrm{M_\odot}$). The final stopping condition is called `none' in the figure, because in these systems, mass transfer ceased and the stars' final masses were too low to reach core carbon burning. These primaries will likely eventually evolve to be white dwarfs, but we stopped their evolution when it was apparent that they had reached this state. 

For a binary to be included in our completed sample, it must eject at least 0.1 solar masses of material and have the primary reach Roche lobe overflow during simulation time. This ensures that the binaries we examine undergo non-conservative mass transfer, the mechanism we are interested in for acting as an enrichment source. Many of the systems at the longest periods did not go through Roche lobe overflow, although some of the higher mass systems beyond our nominal period limit did contribute, confirming that the evolution of a primary star in a binary system is indeed slightly different than that of a single star, even before mass transfer is initiated. 

\section{Results}
\label{sec:results} 

\subsection{Abundances of binary ejecta}

The binary systems in our simulation suite follow the same general evolution as the system described in detail in \citet{deMink2009}: the primary exceeds its Roche lobe radius after core hydrogen exhaustion and begins to transfer material to the secondary. At first this mass transfer is conservative, but the material transfers significant angular momentum to the secondary and it spins up. When it nears its critical rotation rate, it can no longer accept any more material and so any additional material from the primary leaves the system. Typically this happens after relatively small amount of material ($\lesssim 1\msun$) has been added to the secondary. We note that this is the default behaviour in MESA and results in a low accretion efficiency. Measurements of accretion efficiency in post-interaction systems, and also different treatments of mass transfer modelling, suggest higher efficiencies could be possible \citet{2023arXiv231101865M}. Since there is considerable uncertainty, we continue with the default MESA approach but note that our yields are dependent on the amount of material that can remain on the secondary during mass transfer compared to the amount that flows out of the system.

We track how much material is lost from the system, and also the abundance of each element in that material over time. We expect that much of this material will show signatures of hot hydrogen burning, as it will have originated fairly deep in the primary star. The outer, unmodified layers of the primary may have been shed through a normal stellar wind, but mostly will be transferred and accepted by the secondary during the first, conservative, phase of mass transfer. By the time the material starts leaving the system, some or all of that material will be removed from the primary, and the enriched layers will be available. 

\begin{figure*}
    \includegraphics[width=1.8\columnwidth]{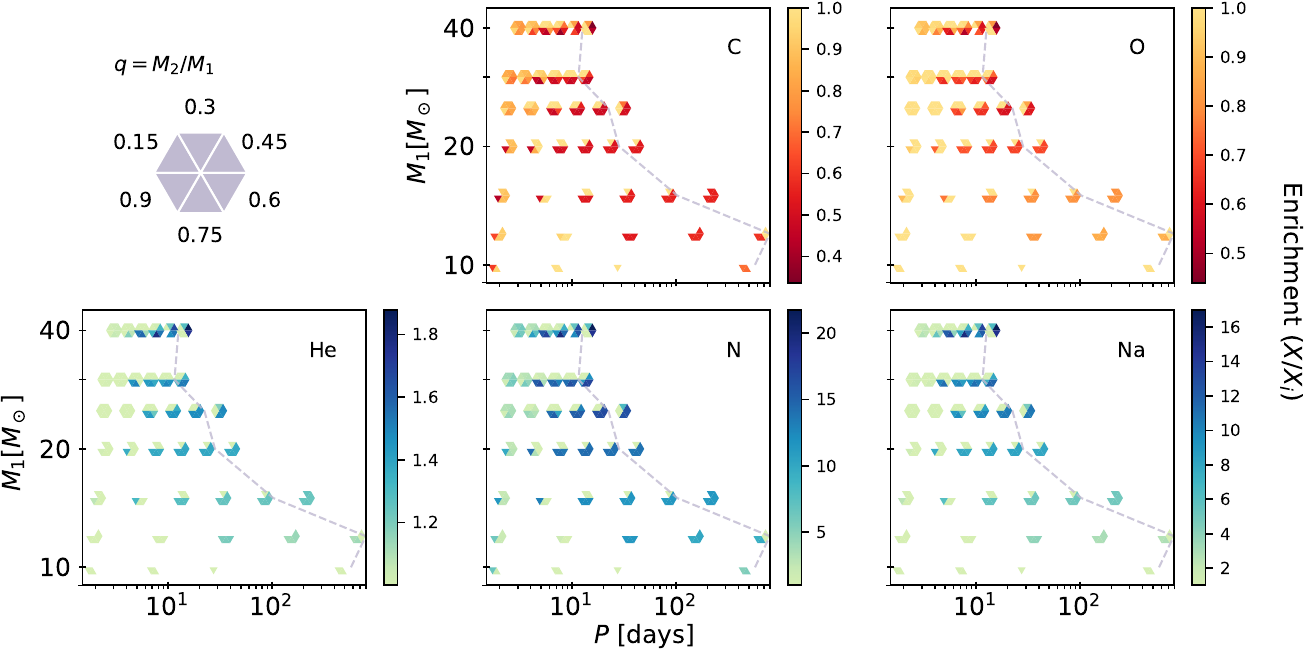}
    \caption{Enrichment of the total ejected material from each simulated binary.    
    The enrichment is the relative change in the mass fraction of the element compared to the initial surface abundances, $X/X_i$.
    Values less than 1 indicate that element is depleted in the ejecta from that binary; values above 1 indicate enhancement.
    The colour indicates the strength of enrichment; darker values correspond to stronger depletion (top panels) or stronger enhancement (bottom panels).
    The strongest signals of enhancement are present at higher $q$, higher $P$, and higher $M_1$.
    }
    \label{fig:enrichment_alpha}
\end{figure*}

For each binary in our simulation suite, we show the abundance of the elements observed to vary in multiple populations (C, O, He, N, Na) in Figure \ref{fig:enrichment_alpha}. Here we show the mass fraction of that element in the ejecta relative to the initial surface abundance of that element, $X/X_i$. 
Points that are darker have ejecta with stronger enrichment signals (associated with more extremely enriched populations) than lighter points. We have a clear signature of hot hydrogen burning, with depleted carbon and oxygen (top panels) and enhanced helium, nitrogen and sodium (bottom panels). Direct comparisons to the observed enhancement values seen in globular clusters are not straightforward since there is a wide variety of enhancement levels between clusters, and some abundances such as helium and often carbon are not measured directly but inferred from photometry or molecular bands. Nevertheless, some clusters have been studied in detail \citep[e.g. the extreme NGC 2808, and the more typical NGC 6752, as described in][]{mult,2019A&ARv..27....8G}. The maximum observed enrichment levels are approximately $X/X_i$=1.5 for helium, up to 100 for nitrogen, and 5 for sodium; oxygen depletion can reach $X/X_i$=0.5, and carbon may be as low as 0.3. 

Systems with high initial primary masses $M_1=40\;\mathrm{M_\odot}$ consistently show the strongest enrichment signals. Enriching binaries (darker points) tend to have higher primary masses $M_1\gtrsim12$, higher $q\gtrsim0.45$ (bottoms and upper right of hexagons), and are at longer periods. These are the systems where most of the pristine outer layers can be accreted onto the secondaries so that it will not be ejected. These periods have also given the primary enough time to allow for significant hydrogen shell burning to proceed, adding to the total amount of hydrogen-processed material that is available. Binaries with shorter periods, particularly the high primary mass systems, do not contribute as much because they reach contact relatively quickly and stop contributing. We also note that there are systems that show essentially no signs of processing with $X/X_i\sim1$. This suggests that non-conservative mass transfer can not only provide enriched gas but can also provide pristine gas for dilution.

\begin{figure}
    \includegraphics[width=\columnwidth]{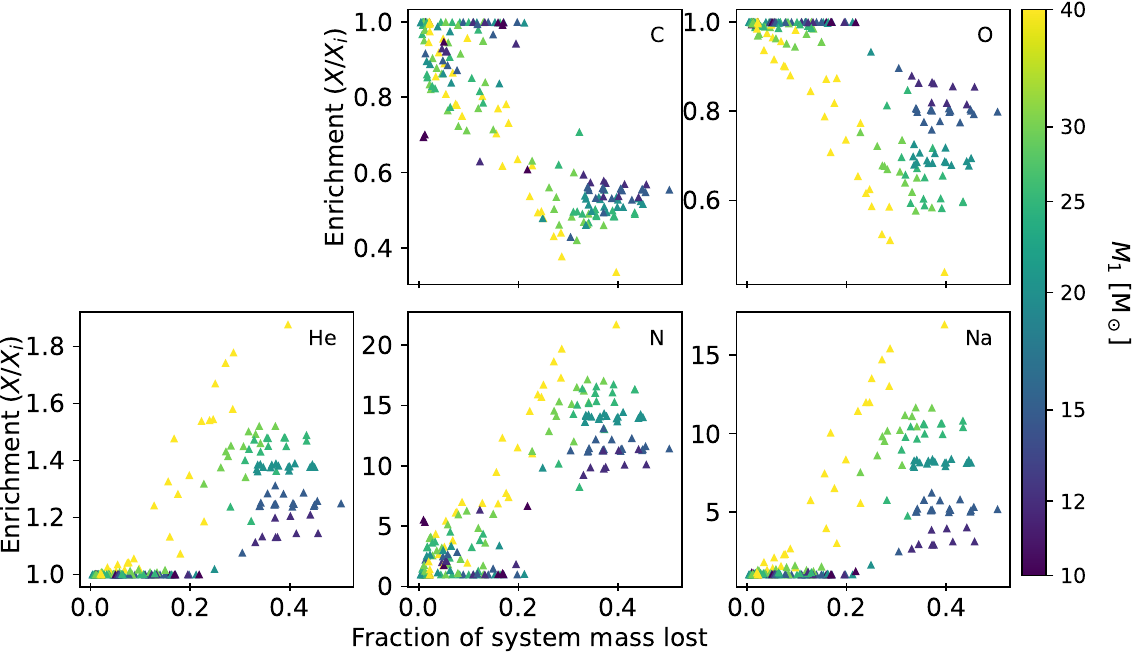}
    \caption{The enrichment, $X/X_i$, in different elements (C, O, He, N, Na) as a function of the amount of mass lost from the system.
    The fraction of system mass lost is $\text{(total mass ejected)}/(M_1+M_2)$.
    Each point corresponds to the yields of a simulated binary in our sample.
    The colour of the points corresponds to the initial primary mass of the binary as shown on the colour bar.}
    \label{fig:enrichment_frac_m1}
\end{figure}

The total amount of mass ejected by each system varies considerably, as shown in Figure \ref{fig:enrichment_frac_m1}. Here we show the enrichment in five representative elements as a function of the fraction of the initial system mass that has been ejected from the system. In general, less enriching binaries with $X/X_i\sim1$ eject a smaller fraction of their mass. A transition appears when around 20\% of the system mass has been ejected from the system. We also see that the maximum fraction of system mass lost is under 50\% of the total mass of each binary system. 

\begin{table}
    \begin{tabular}{c|c|c|c}
Element & Minimum &  Maximum & Median \\
\hline
H & 0.71 &  1.00 &  0.99 \\
He & 1.00 & 1.88 &  1.02 \\
Li & 0.01 & 0.68 & 0.02 \\
Be & 0.02 &  1.00 &  0.15 \\
B & 0.02 &  0.998 &  0.14 \\
C & 0.34 & 1.0 & 0.77 \\
N & 1.00 &  21.72 & 5.76 \\
O & 0.44 &  1.0 &  0.96 \\
F & 0.35 &  1.001 &  0.95 \\
Na & 1.00 &  16.97 &  1.49 \\
Mg & 0.96 &  1.0 &  0.999 \\
Al & 1.00 &  2.21 &  1.04 \\
Si & 1.00 &  1.00 &  1.00 \\
Ca & 1.00 &  1.00 &  1.00 \\
Ti & 1.00 &  1.00 &  1.00 \\
Fe & 1.00 &  1.00 &  1.00 \\
Ni & 1.00 &  1.00 & 1.00 \\
    \end{tabular}
    \caption{Maximum, minimum, and median enrichment/depletion relative to the initial mass fraction of each element, $X/X_i$, in the ejecta of the suite of simulated binaries. A value above 1 denotes enrichment, and below 1 denotes depletion.
    \label{tab:binary_all_trends}}
\end{table}

 Our simulations followed many more elements than just those shown in the previous figures. The main ones are listed in Table \ref{tab:binary_all_trends}, where we give the minimum and maximum enrichment or depletion $X/X_i$ in our suite of 204 systems, and the median $X/X_i$ of all systems. The iron peak elements (Ti, Fe, and Ni) do not change, and neither do the heavier alpha elements (Ca, Si). 

The fragile light elements Li, Be, and B are all significantly depleted. In some systems, hydrogen is depleted which indicates  the ejection of H-burned, He-enhanced material, but the overall depletion of hydrogen is very small. The other light elements show the expected  signatures of multiple populations (and hot hydrogen burning): depletion in C, O, and Mg and enhancement in He, N, Na, and Al. We have also confirmed that the expected correlations (e.g. C-O, N-Na) and anti-correlations (e.g. C-N, Na-O) are found in the ejecta from massive interacting binaries. Since we are essentially probing the abundances of hot-hydrogen-processed material, this is the expected outcome of our choices of binary parameters. 

We make special note of the Mg-Al anti-correlation, since that is seen in only the most massive globular clusters, and is often used as an important diagnostic to test models of multiple population yields. Most of our binaries do not show significant change in their magnesium abundance. If magnesium does change, we see a very tight anti-correlation between Al and Mg. The maximum depletion of [Mg/Fe] is approximately 0.1 dex (with a corresponding enhancement in [Al/Fe] of 0.3 dex). The systems with primary masses of 35 and 40 $\msun$ are responsible for almost all the change in Mg, as the Mg-Al cycle requires very high temperatures to be significantly activated. 

\subsection{Timescales for binary ejecta}

\begin{figure}
    \includegraphics[width=1.1\columnwidth]{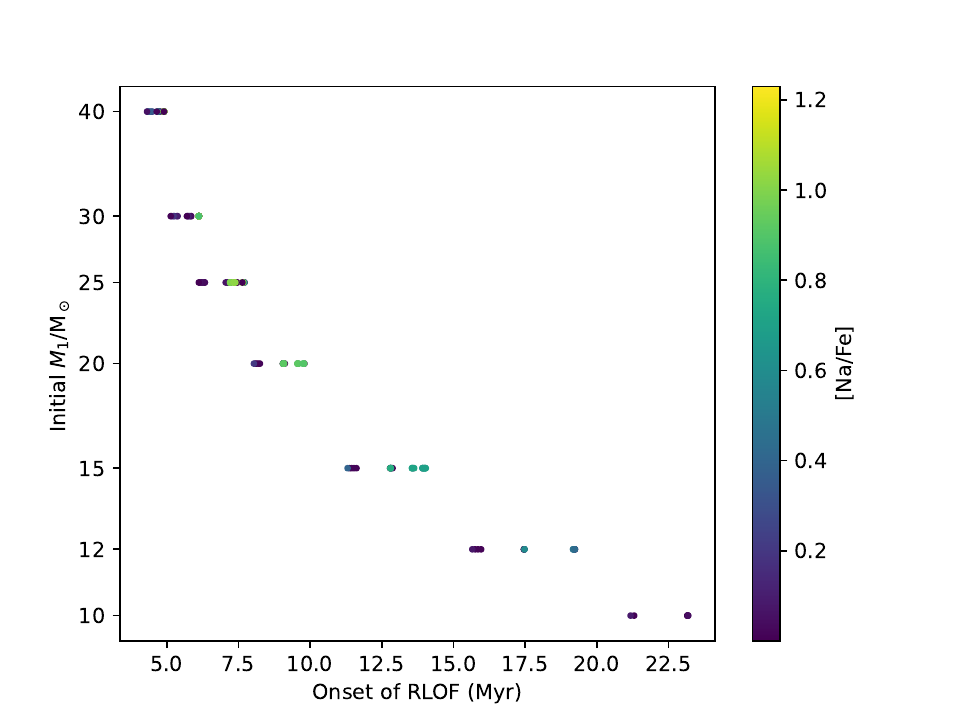}
    \caption{The time of onset of Roche lobe overflow as a function of initial primary mass. Systems with longer initial periods begin mass transfer at later times for a given mass.
    [Na/Fe] is shown on the colour bar as an indication of the enrichment level of the binary ejecta.}
    \label{fig:rlo_init}
\end{figure}

The timescale on which these systems eject mass is an important factor in their ability to act as an enrichment source. In all cases, the duration of the mass transfer event is short ($\lesssim0.5$ Myr, and the non-conservative portion of the mass transfer event is half or less of that time), compared to the evolution of the system before the onset of mass transfer. Therefore, we assign a single time of mass ejection for all of the binaries in our samples, shown in Figure \ref{fig:rlo_init}.  We define this time as the age of the system when the primary's radius first exceeds the Roche lobe radius. The lowest mass primaries eject on timescales $\gtrsim23$~Myr, and the highest mass primaries we simulate can eject their material as early as $\sim4.3$~Myr. The time of mass transfer onset increases with increasing period, but this results in less spread ($\sim5$~Myr for our lowest mass systems than the difference caused by the evolutionary timescale of the primary. 

\subsection{Expected yields from a population of binaries} \label{sec:expected}

 The total amount of available material, and its composition, to create the enriched population in star clusters will not come from a single binary, but instead from the ensemble of the binary population.  Now that we have the nucleosynthetic yields for individual systems, we can combine them by weighting each binary by its likelihood of being part of the population. To do this, we need to make some assumptions about the properties of the binary population of a forming cluster. For this work, we make the simplifying assumption that the binary population in a cluster is the same as the field population. Binary populations can be modified dynamically through interactions with other stars and the natal gas in forming clusters \citep{2021MNRAS.501.4464C}, but the initial distribution cannot be very different from the field population. We use the same binary sampling routine as in \citet{2021MNRAS.501.4464C}, which is based on the field binary properties from \citet{2017ApJS..230...15M}. The binary fraction and period distribution are functions of the primary mass, and the mass ratio distribution depends on both primary mass and period.
 
We determine the probability of having a binary in a three-dimensional grid covering primary mass, mass ratio, and period.  We do not consider the eccentricity distribution of the binaries as we only simulate circular binaries. We interpolate the yields of each element and the total amount of mass ejected from our binary system simulations onto the gridpoints of primary mass, mass ratio and period. We do not extrapolate any yields beyond the boundaries of our simulations ($1.87<P<661.65$~days, $10<M_1<40\;\mathrm{M_\odot}$, and $0.15<q<0.9$ in steps of $\Delta (\log P)=0.01$~[days], $\Delta M_1=0.1\;\mathrm{M_\odot}$, and $\Delta q =0.01$) but we do calculate the probabilities over the full binary population explored in \citet{2021MNRAS.501.4464C} which covers masses from 0.08 to 150 $\mathrm{M_\odot}$, periods between $10^{0.5}$ and $10^{7.5}$ days, and all mass ratios.  The total yield of each element is then the sum, over all grid points, of the individual binary yield weighted by the probability of having such a binary. We are not considering any material outside the boundaries of our simulations, so the results presented here should be considered as the yields from hot hydrogen burning only. 

\begin{table}
    \centering
    \begin{tabular}{c|c|c}
Element & $X/X_i$ &  [X/Fe] \\
\hline
He & 1.335 & Y=0.335  \\
C & 0.558 & 0.326  \\
N & 12.048 &  1.079  \\
O & 0.748 &  0.313  \\
Na & 7.169 &  0.855  \\
Mg & 0.985 &  0.342  \\
Al & 1.431 &  0.155 \\
    \end{tabular}
    \caption{Total enrichment or depletion relative to the initial mass fraction of each element, $X/X_i$, and also the new value of Y or [X/Fe], from a population of binary stars weighted by their likelihood of occurring in the field.
    \label{tab:population}}
\end{table}

The results for the elements relevant for multiple populations are given in Table \ref{tab:population} in two different ways: the total enrichment/depletion relative to the initial mass fraction of the element, $X/X_i$, and also the resultant [X/Fe] value (except for helium which is reported as the new value of Y). This weighted average of the individual binary yields is precisely what is expected to explain the multiple populations problem: enhancement in helium up to above Y=0.3, enhancement in oxygen, sodium, and aluminum accompanied by depletion in carbon, nitrogen, and magnesium. The levels of enhancement or depletion (e.g. large values of nitrogen, small values of magnesium) are also what is seen in detailed spectroscopic observations in globular clusters \citep[e.g.][]{2019A&ARv..27....8G}. 

Using the same weighting of the individual binaries, we calculate that the population of binary stars will release 23\% of its total mass in ejecta with this composition. Earlier work investigating the binary ejection hypothesis \citep{2019MNRAS.486.1146H} assumed 3 possible values for this quantity (10, 25, or 50\%) and found that 25\% was the best match to observations, so we can confirm this was a reasonable assumption. Finally, we calculate the binary-weighted average timescale for this ejecta to be released to be just under 12 Myr. For comparison, the same population of stars as our binaries, but evolved as two single stars, would eject approximately 3.8\% of their mass on a timescale of 14 Myr, a reduction of about 1/6 in total mass compared to the binary population.  



\section{Summary and discussion}

Using the stellar evolution code MESA, we simulated the simultaneous evolution of both stars in 204 binary systems with parameters that covered the range over which we expected to see non-conservative case B mass transfer: $10\leq M_1\leq40\msun$, $0.15\leq q\leq 0.9$, and $2\lesssim P\lesssim700$ days. We tracked the composition of the material which is lost from these systems and compared the abundances to the observed and inferred abundances of multiple populations in globular clusters. The yields of our binaries showed the expected signature of hot hydrogen burning: they are enhanced in helium, nitrogen, sodium, and aluminium, and depleted in carbon, oxygen, and magnesium. Fragile light elements Li, Be, and B are destroyed, and the iron peak elements are unchanged. The material is ejected between when the stars are between 4 and 23 Myr old, depending on the mass of the primary. 

We then combined these yields with our expectation for the distribution of these binaries in mass, mass ratio, and period space to calculate the total yield of a normal, field-like population of binaries. We find that the total population will release approximately 25\% of its total mass in ejecta on a typical timescale of 12 Myr, and with overall abundances which match the most extreme populations in massive globular clusters. Dilution of this material with pristine gas will provide the range of abundances seen in clusters. This confirms the suggestion of \citet{deMink2009} that a population of massive binaries are a viable source of enriched material out of which to form multiple populations. 

The total mass ejected is consistent with the amount needed to produce multiple populations in the context of pollution during cluster assembly \citep[e.g.][]{2019MNRAS.486.1146H}. If instead we assume that clusters self-enrich some time after all the un-enriched stars have formed, then binaries do not produce enough material to solve the mass budget problem. However, most other stellar populations, individually or even when considering multiple enrichment sources \citep{2010MNRAS.407..277S}, are also not able to create enough enriched material. This problem, combined with the increasing evidence of no age difference between the different populations, points strongly toward an enrichment source which must be acting during cluster assembly, and enriching only some of the material out of which the cluster will assemble during that process. 

If indeed enrichment must happen during cluster assembly, then the timescale to produce the enriched material must be shorter than the timescale on which cluster stars are forming. We find that binaries will produce their enriched material within about 12 Myr. Nearby young clusters have observed age spreads of at most a few Myr \citep[see][for studies of R136 and Orion respectively]{1998ApJ...493..180M,2010ApJ...722.1092D}. Such a short formation time would pose a problem for the massive binaries to release their enriched material in time to be relevant. On the other hand, simulations of massive star cluster formation suggest that star formation durations up to 20 Myr \citep{2024A&A...681A..28A} are possible, with many simulations suggesting that about 10 Myr is needed to assemble massive clusters \citep[e.g.][]{2018NatAs...2..725H,10.1093/mnras/stad1572}. Dating very young stars is notoriously difficult, and simulations include different prescriptions for important physical processes. Efforts will be needed on both fronts to resolve these discrepancies. We also note that the most enriching systems are those with the highest primary masses, which evolve on shorter timescales and so may do most of the necessary enrichment on timescales which are as short as about 4 Myr. 

In this paper, we have assumed that the binaries are evolving in isolation, and that the overall population is the same as that seen in the field. However, binaries in clusters are dynamically modified \citep{1975MNRAS.173..729H}, even during the very earliest stages of cluster formation \citep{2021MNRAS.501.4464C}. Dynamical evolution, in general, reduces binary periods and increases binary mass ratios. All our binaries with mass ratios closer to 1 produce more enriched material than those with lower $q$. In general, a longer binary period is more conducive to producing more enriched material, mainly because a longer period allowed the primary time to process more of its material though hot hydrogen burning in its hydrogen-burning shell. How dynamics would affect the abundances depends on when the system's period is reduced. Some systems with periods beyond our current upper period boundary would be pushed into a region where they would now start contributing; others may have their period reduced early and may produce less enriched ejecta; but some are likely to have their periods reduced after most of the burning has happened, and then they might eject even more material than if they had stayed at their original period. If we had the complete dynamical evolution of a population of binaries inside a forming cluster, it may be possible to extrapolate along the binary tracks calculate in this paper to determine the final properties of a modified, non-field binary population. 

Massive binaries provide a very attractive solution to the source of multiple populations in globular clusters, simply because they are so common and so commonplace. Massive stars are formed in clusters. Massive stars are formed in binaries and higher order systems such as triples or quadruples. Multiple star systems transfer mass and undergo non-conservative mass transfer. They are dynamically modified in a cluster environment. We have shown that a population of massive binaries will produce a significant amount of material with the right abundances, up to and including the Al-Mg anti-correlation. They eject this material on a timescale which is comparable to the assembly timescale of massive star clusters. Just as any models of star cluster formation should include binary stars, any model of multiple populations should include binary stars. It remains to be seen whether binaries alone can produce all the signatures of multiple populations, but since we know that binaries exist in significant numbers, we must include them in our models. 

\vspace{0.18cm}

This work is supported by the Natural Sciences and Engineering Research Council of Canada. This research was enabled in part by support provided by Compute Ontario (https://www.computeontario.ca) and the Digital Research Alliance of Canada (https://alliancecan.ca).

\bibliography{ref}{}

\begin{thebibliography}{}
\expandafter\ifx\csname natexlab\endcsname\relax\def\natexlab#1{#1}\fi
\providecommand{\url}[1]{\href{#1}{#1}}
\providecommand{\dodoi}[1]{doi:~\href{http://doi.org/#1}{\nolinkurl{#1}}}
\providecommand{\doeprint}[1]{\href{http://ascl.net/#1}{\nolinkurl{http://ascl.net/#1}}}
\providecommand{\doarXiv}[1]{\href{https://arxiv.org/abs/#1}{\nolinkurl{https://arxiv.org/abs/#1}}}

\bibitem[{{Adamo} {et~al.}(2023){Adamo}, {Usher}, {Pfeffer}, \& {Claeyssens}}]{2023MNRAS.525L...6A}
{Adamo}, A., {Usher}, C., {Pfeffer}, J., \& {Claeyssens}, A. 2023, \mnras, 525, L6, \dodoi{10.1093/mnrasl/slad084}

\bibitem[{{Andersson} {et~al.}(2024){Andersson}, {Mac Low}, {Agertz}, {Renaud}, \& {Li}}]{2024A&A...681A..28A}
{Andersson}, E.~P., {Mac Low}, M.-M., {Agertz}, O., {Renaud}, F., \& {Li}, H. 2024, \aap, 681, A28, \dodoi{10.1051/0004-6361/202347792}

\bibitem[{{Bastian} \& {Lardo}(2018)}]{mult}
{Bastian}, N., \& {Lardo}, C. 2018, \araa, 56, 83, \dodoi{10.1146/annurev-astro-081817-051839}

\bibitem[{{Choi} {et~al.}(2016){Choi}, {Dotter}, {Conroy}, {Cantiello}, {Paxton}, \& {Johnson}}]{Choi2016}
{Choi}, J., {Dotter}, A., {Conroy}, C., {et~al.} 2016, \apj, 823, 102, \dodoi{10.3847/0004-637X/823/2/102}

\bibitem[{{Cournoyer-Cloutier} {et~al.}(2021){Cournoyer-Cloutier}, {Tran}, {Lewis}, {Wall}, {Harris}, {Mac Low}, {McMillan}, {Portegies Zwart}, \& {Sills}}]{2021MNRAS.501.4464C}
{Cournoyer-Cloutier}, C., {Tran}, A., {Lewis}, S., {et~al.} 2021, \mnras, 501, 4464, \dodoi{10.1093/mnras/staa3902}

\bibitem[{{Da Rio} {et~al.}(2010){Da Rio}, {Robberto}, {Soderblom}, {Panagia}, {Hillenbrand}, {Palla}, \& {Stassun}}]{2010ApJ...722.1092D}
{Da Rio}, N., {Robberto}, M., {Soderblom}, D.~R., {et~al.} 2010, \apj, 722, 1092, \dodoi{10.1088/0004-637X/722/2/1092}

\bibitem[{de~Mink {et~al.}(2009)de~Mink, Pols, Langer, \& Izzard}]{deMink2009}
de~Mink, S., Pols, O., Langer, N., \& Izzard, R. 2009, \aap, 507

\bibitem[{{Decressin} {et~al.}(2007){Decressin}, {Meynet}, {Charbonnel}, {Prantzos}, \& {Ekstr{\"o}m}}]{frms}
{Decressin}, T., {Meynet}, G., {Charbonnel}, C., {Prantzos}, N., \& {Ekstr{\"o}m}, S. 2007, \aap, 464, 1029, \dodoi{10.1051/0004-6361:20066013}

\bibitem[{{Dickson} {et~al.}(2023){Dickson}, {H{\'e}nault-Brunet}, {Baumgardt}, {Gieles}, \& {Smith}}]{2023MNRAS.522.5320D}
{Dickson}, N., {H{\'e}nault-Brunet}, V., {Baumgardt}, H., {Gieles}, M., \& {Smith}, P.~J. 2023, \mnras, 522, 5320, \dodoi{10.1093/mnras/stad1254}

\bibitem[{{Eggleton}(1983)}]{Eggleton1983}
{Eggleton}, P.~P. 1983, \apj, 268, 368, \dodoi{10.1086/160960}

\bibitem[{{Gieles} {et~al.}(2018){Gieles}, {Charbonnel}, {Krause}, {H{\'e}nault-Brunet}, {Agertz}, {Lamers}, {Bastian}, {Gualandris}, {Zocchi}, \& {Petts}}]{2018MNRAS.478.2461G}
{Gieles}, M., {Charbonnel}, C., {Krause}, M. G.~H., {et~al.} 2018, \mnras, 478, 2461, \dodoi{10.1093/mnras/sty1059}

\bibitem[{{Glebbeek} {et~al.}(2009){Glebbeek}, {Gaburov}, {de Mink}, {Pols}, \& {Portegies Zwart}}]{dutch1}
{Glebbeek}, E., {Gaburov}, E., {de Mink}, S.~E., {Pols}, O.~R., \& {Portegies Zwart}, S.~F. 2009, \aap, 497, 255, \dodoi{10.1051/0004-6361/200810425}

\bibitem[{{Gratton} {et~al.}(2019){Gratton}, {Bragaglia}, {Carretta}, {D'Orazi}, {Lucatello}, \& {Sollima}}]{2019A&ARv..27....8G}
{Gratton}, R., {Bragaglia}, A., {Carretta}, E., {et~al.} 2019, \aapr, 27, 8, \dodoi{10.1007/s00159-019-0119-3}

\bibitem[{{Heggie}(1975)}]{1975MNRAS.173..729H}
{Heggie}, D.~C. 1975, \mnras, 173, 729, \dodoi{10.1093/mnras/173.3.729}

\bibitem[{{Henneco} {et~al.}(2024){Henneco}, {Schneider}, \& {Laplace}}]{2024A&A...682A.169H}
{Henneco}, J., {Schneider}, F.~R.~N., \& {Laplace}, E. 2024, \aap, 682, A169, \dodoi{10.1051/0004-6361/202347893}

\bibitem[{{Howard} {et~al.}(2018){Howard}, {Pudritz}, \& {Harris}}]{2018NatAs...2..725H}
{Howard}, C.~S., {Pudritz}, R.~E., \& {Harris}, W.~E. 2018, Nature Astronomy, 2, 725, \dodoi{10.1038/s41550-018-0506-0}

\bibitem[{{Howard} {et~al.}(2019){Howard}, {Pudritz}, {Sills}, \& {Harris}}]{2019MNRAS.486.1146H}
{Howard}, C.~S., {Pudritz}, R.~E., {Sills}, A., \& {Harris}, W.~E. 2019, \mnras, 486, 1146, \dodoi{10.1093/mnras/stz924}

\bibitem[{{Jermyn} {et~al.}(2023){Jermyn}, {Bauer}, {Schwab}, {Farmer}, {Ball}, {Bellinger}, {Dotter}, {Joyce}, {Marchant}, {Mombarg}, {Wolf}, {Sunny Wong}, {Cinquegrana}, {Farrell}, {Smolec}, {Thoul}, {Cantiello}, {Herwig}, {Toloza}, {Bildsten}, {Townsend}, \& {Timmes}}]{jermyn2023}
{Jermyn}, A.~S., {Bauer}, E.~B., {Schwab}, J., {et~al.} 2023, \apjs, 265, 15, \dodoi{10.3847/1538-4365/acae8d}

\bibitem[{{Krause} {et~al.}(2020){Krause}, {Offner}, {Charbonnel}, {Gieles}, {Klessen}, {V{\'a}zquez-Semadeni}, {Ballesteros-Paredes}, {Girichidis}, {Kruijssen}, {Ward}, \& {Zinnecker}}]{2020SSRv..216...64K}
{Krause}, M. G.~H., {Offner}, S. S.~R., {Charbonnel}, C., {et~al.} 2020, \ssr, 216, 64, \dodoi{10.1007/s11214-020-00689-4}

\bibitem[{{Marchant} \& {Bodensteiner}(2023)}]{2023arXiv231101865M}
{Marchant}, P., \& {Bodensteiner}, J. 2023, arXiv e-prints, arXiv:2311.01865, \dodoi{10.48550/arXiv.2311.01865}

\bibitem[{{Martell} {et~al.}(2011){Martell}, {Smolinski}, {Beers}, \& {Grebel}}]{2011A&A...534A.136M}
{Martell}, S.~L., {Smolinski}, J.~P., {Beers}, T.~C., \& {Grebel}, E.~K. 2011, \aap, 534, A136, \dodoi{10.1051/0004-6361/201117644}

\bibitem[{{Massey} \& {Hunter}(1998)}]{1998ApJ...493..180M}
{Massey}, P., \& {Hunter}, D.~A. 1998, \apj, 493, 180, \dodoi{10.1086/305126}

\bibitem[{{Milone} \& {Marino}(2022)}]{2022Univ....8..359M}
{Milone}, A.~P., \& {Marino}, A.~F. 2022, Universe, 8, 359, \dodoi{10.3390/universe8070359}

\bibitem[{{Moe} \& {Di Stefano}(2017)}]{2017ApJS..230...15M}
{Moe}, M., \& {Di Stefano}, R. 2017, \apjs, 230, 15, \dodoi{10.3847/1538-4365/aa6fb6}

\bibitem[{{Nguyen}(2023)}]{NguyenThesis}
{Nguyen}, M. 2023, Master's thesis, McMaster University.
\newblock \url{{http://hdl.handle.net/11375/28969}}

\bibitem[{{Offner} {et~al.}(2023){Offner}, {Moe}, {Kratter}, {Sadavoy}, {Jensen}, \& {Tobin}}]{2023ASPC..534..275O}
{Offner}, S.~S.~R., {Moe}, M., {Kratter}, K.~M., {et~al.} 2023, in Astronomical Society of the Pacific Conference Series, Vol. 534, Astronomical Society of the Pacific Conference Series, ed. S.~{Inutsuka}, Y.~{Aikawa}, T.~{Muto}, K.~{Tomida}, \& M.~{Tamura}, 275

\bibitem[{{Paxton} {et~al.}(2011){Paxton}, {Bildsten}, {Dotter}, {Herwig}, {Lesaffre}, \& {Timmes}}]{Paxton2011}
{Paxton}, B., {Bildsten}, L., {Dotter}, A., {et~al.} 2011, \apjs, 192, 3, \dodoi{10.1088/0067-0049/192/1/3}

\bibitem[{{Paxton} {et~al.}(2013){Paxton}, {Cantiello}, {Arras}, {Bildsten}, {Brown}, {Dotter}, {Mankovich}, {Montgomery}, {Stello}, {Timmes}, \& {Townsend}}]{Paxton2013}
{Paxton}, B., {Cantiello}, M., {Arras}, P., {et~al.} 2013, \apjs, 208, 4, \dodoi{10.1088/0067-0049/208/1/4}

\bibitem[{{Paxton} {et~al.}(2015){Paxton}, {Marchant}, {Schwab}, {Bauer}, {Bildsten}, {Cantiello}, {Dessart}, {Farmer}, {Hu}, {Langer}, {Townsend}, {Townsley}, \& {Timmes}}]{Paxton2015}
{Paxton}, B., {Marchant}, P., {Schwab}, J., {et~al.} 2015, \apjs, 220, 15, \dodoi{10.1088/0067-0049/220/1/15}

\bibitem[{{Paxton} {et~al.}(2018){Paxton}, {Schwab}, {Bauer}, {Bildsten}, {Blinnikov}, {Duffell}, {Farmer}, {Goldberg}, {Marchant}, {Sorokina}, {Thoul}, {Townsend}, \& {Timmes}}]{Paxton2018}
{Paxton}, B., {Schwab}, J., {Bauer}, E.~B., {et~al.} 2018, \apjs, 234, 34, \dodoi{10.3847/1538-4365/aaa5a8}

\bibitem[{{Paxton} {et~al.}(2019){Paxton}, {Smolec}, {Schwab}, {Gautschy}, {Bildsten}, {Cantiello}, {Dotter}, {Farmer}, {Goldberg}, {Jermyn}, {Kanbur}, {Marchant}, {Thoul}, {Townsend}, {Wolf}, {Zhang}, \& {Timmes}}]{Paxton2019}
{Paxton}, B., {Smolec}, R., {Schwab}, J., {et~al.} 2019, \apjs, 243, 10, \dodoi{10.3847/1538-4365/ab2241}

\bibitem[{{Reddy} {et~al.}(2006){Reddy}, {Lambert}, \& {Allende Prieto}}]{reddy2006}
{Reddy}, B.~E., {Lambert}, D.~L., \& {Allende Prieto}, C. 2006, \mnras, 367, 1329, \dodoi{10.1111/j.1365-2966.2006.10148.x}

\bibitem[{{Ritter}(1988)}]{Ritter1988}
{Ritter}, H. 1988, \aap, 202, 93

\bibitem[{Rizzuti {et~al.}(2023)Rizzuti, Hirschi, Arnett, Georgy, Meakin, Murphy, Rauscher, \& Varma}]{10.1093/mnras/stad1572}
Rizzuti, F., Hirschi, R., Arnett, W.~D., {et~al.} 2023, Monthly Notices of the Royal Astronomical Society, 523, 2317, \dodoi{10.1093/mnras/stad1572}

\bibitem[{{Sana} {et~al.}(2012){Sana}, {de Mink}, {de Koter}, {Langer}, {Evans}, {Gieles}, {Gosset}, {Izzard}, {Le Bouquin}, \& {Schneider}}]{2012Sci...337..444S}
{Sana}, H., {de Mink}, S.~E., {de Koter}, A., {et~al.} 2012, Science, 337, 444, \dodoi{10.1126/science.1223344}

\bibitem[{{Sills} \& {Glebbeek}(2010)}]{2010MNRAS.407..277S}
{Sills}, A., \& {Glebbeek}, E. 2010, \mnras, 407, 277, \dodoi{10.1111/j.1365-2966.2010.16876.x}

\bibitem[{{Ventura} \& {D'Antona}(2009)}]{2009A&A...499..835V}
{Ventura}, P., \& {D'Antona}, F. 2009, \aap, 499, 835, \dodoi{10.1051/0004-6361/200811139}

\bibitem[{{Ying} {et~al.}(2023){Ying}, {Chaboyer}, {Boudreaux}, {Slaughter}, {Boylan-Kolchin}, \& {Weisz}}]{2023AJ....166...18Y}
{Ying}, J.~M., {Chaboyer}, B., {Boudreaux}, E.~M., {et~al.} 2023, \aj, 166, 18, \dodoi{10.3847/1538-3881/acd9b1}

\end{thebibliography}
\bibliographystyle{aasjournal}

\end{document}